\documentclass[usegraphicx,usenatbib]{mn2e}
\usepackage{psfig}
\usepackage{pslatex}

\renewcommand\[{\begin{equation}}
\renewcommand\]{\end{equation}}
\def\Mpc{\,{\rm Mpc}}\def\kpc{\,{\rm kpc}}
\def\cm{\,{\rm cm}}\def\erg{\,{\rm erg}}\def\g{\,{\rm g}}\def\s{\,{\rm s}}
\def\keV{\,{\rm keV}}
\def\Myr{\,{\rm Myr}}\def\Gyr{\,{\rm Gyr}}
\def\K{\,{\rm K}}
\def\msun{h^{-1}\,{\rm M_{\odot}}}
\def\kms{\,{\rm km s}^{-1}}

\def\kB{k_{\rm B}}
\def\d{{\rm d}}\def\e{{\rm e}}

\def\i{\relax\ifmmode{\rm i}\else\char16\fi}
\def\lesssim{{_ <\atop{^\sim}}}

\def\grtsim{{_ >\atop{^\sim}}}
\def\gta{\grtsim}

\def\fracj#1#2{{\textstyle{#1\over#2}}}

\def\lesssim{\mathrel{\hbox{\rlap{\hbox{\lower4pt\hbox{$\sim$}}}\hbox{$<$}}}}
\def\gtrsim{\mathrel{\hbox{\rlap{\hbox{\lower4pt\hbox{$\sim$}}}\hbox{$>$}}}}

\def\apj#1 #2 {ApJ, {\bf #1}, #2}
\def\aj#1 #2 {AJ, {\bf #1}, #2}
\def\mn#1 #2 {MNRAS, {\bf #1}, #2}
\def\aa#1 #2 {A\&A, {\bf #1}, #2}

\begin{document}

   \title[Simple models of cooling flows]
   {Simple models of cooling flows}

   \author[Kaiser \& Binney]
          {Christian R. Kaiser$^1$ \& James Binney$^2$
           \\
		   $^1$ Department of Physics \& Astronomy, University of Southampton, Southampton, SO17 1BJ \\
           $^2$Theoretical Physics, 1 Keble Road, Oxford OX1 3NP\\
          }

   \date{Received ...; accepted ...}

   \maketitle

\begin{abstract}
A semi-analytic model of cluster cooling flows is presented. The model
assumes that episodic nuclear activity followed by radiative cooling
without mass-dropout cycles the cluster gas between a relatively
homogeneous, nearly isothermal post-outburst state and a cuspy
configuration in which a cooling catastrophe initiates the next
nuclear outburst. Fitting the model to {\it Chandra\/} data for the
Hydra cluster, a lower limit of $284\Myr$ until the next outburst of
Hydra A is derived. Density, temperature and emission-measure profiles
at several times prior to the cooling catastrophe are presented. It
proves possible to fit the mass $M(\sigma)$ with entropy index
$P\rho^{-\gamma}$ less than $\sigma$ to a simple power-law form, which
is almost invariant as the cluster cools. We show that radiative
cooling automatically establishes this power-law form if the entropy
index was constant throughout the cluster gas at some early epoch or
after an AGN activity cycle. To high precision, the central value of
$\sigma$ decreases linearly in time. The fraction of clusters in a
magnitude-limited sample that have gas cooler than $T$ is calculated,
and is shown to be small for $T=2\keV$. Similarly, only 1 percent of
clusters in such a sample contain gas with $P\rho^{-\gamma} <
2\keV\cm^2$. Entropy production in shocks is shown to be small. The
entropy that is radiated from the cluster can be replaced if a few
percent of the cluster gas passes through bubbles heated during an
outburst of the AGN.

\end{abstract}


\section{Introduction}
 Observations from the {\it Chandra\/} and {\it XMM-Newton\/} missions
have shown that the intergalactic media of cooling-flow clusters is
not a multiphase medium of the type required by the theory of cooling
flows that has been widely accepted for over a decade
\citep[e.g.][]{af94}. Moreover, when these data are taken together
with earlier radio and optical data, it is now clear that over the
lifetimes of these systems very little gas has cooled to temperatures
much below the virial temperature.  Consequently, these systems cannot
be the scenes of a quiescent steady inflow regulated by distributed
`mass dropout' as that theory postulated. In the light of the new
observations \citep{bbk01,mp01,mwn01,mkp02,mbf02}, increasing
numbers of workers \citep[e.g.][and references therein]{bmc02} are
accepting the view that catastrophic cooling of intergalactic gas to
very low temperatures takes place only at very small radii, and that
on larger scales the intergalactic medium is periodically reheated by
an active galactic nucleus (AGN) that is fed by central mass
dropout. This general picture has been argued for several times over
the last decade \citep{tb93,bt95,jb96,jb99,co97,jb01}, but the physics
involved is complex because the problem is inherently
three-dimensional and unsteady, and many details remain to be filled
in now that clearer observations are becoming available.

Recently, a number of authors have re-investigated the possibility
that heat transported from the cluster outskirts to the centre by
thermal conduction could contribute significantly to re-plenishing
radiated energy \citep{vsf02,fvm02,zn02,rb02}. The classical argument
against thermal conduction remains that there is a narrow range of
conductivity in which conduction is numerically significant but fails
to eliminate the cooling region entirely
\citep{BinneyC,MeiksCf}. Moreover, conductivities that predict
significant conductive heat input to dense regions now, are probably
incompatible with the existence of cooling-flow clusters since they
would have caused cluster gas to evaporate early in the universe
\citep{al02}. In this paper we model the effects of AGN alone, and
neglect thermal conduction.

Although the mechanism by which the AGN heats the IGM is unclear --
possibilities include the impact on ambient thermal plasma of
collimated outflows from the AGN \citep{hrb98,ka98b,jb99} and inverse
Compton scattering by thermal plasma of hard photons from the AGN
\citep{co97,co01} -- features have been observed near the centres of
several clusters that are almost certainly bubbles of hot, low-density
plasma that have been created by the AGN.  Simulations
\citep{cbkbf00,ssb01,bk01,qbb01,bkc01} suggest that these rise through
the cluster's gravitational potential well on a dynamical
timescale. The details of how a rising bubble mixes with surrounding
gas and disperses its energy around the intracluster medium cannot be
securely deduced from simulations, because they depend on what happens
on very small scales at the edge of a bubble. The general idea that
heating by AGN results in a kind of convection in the ICM is probably
correct, however. The purpose of this note is to present a simple
semi-analytical model of the life-cycle of a typical cluster that this
process drives. 

Section 2 defines the model, fits  its initial condition to {\it Chandra\/} 
data for the Hydra cluster, and shows its observable characteristics at a
number of later epochs. Section 3 discusses the creation of
entropy during an outburst of the AGN and estimates the fraction of the
cluster gas that an outburst processes  through bubbles. Section 4
calculates the distribution of cluster-centre temperatures in a
magnitude-limited sample of clusters and shows that temperatures below $\sim2\keV$ will
very rarely be detected. Section 5 similarly calculates the distribution of
the minimum values of the specific entropy that will be detected in a
survey. Section 6 sums up.
Throughout we assume a flat cosmology with
$\Omega_\Lambda=0.7$ and $H_0=65\kms\Mpc^{-1}$.

\begin{figure}
\centerline{\psfig{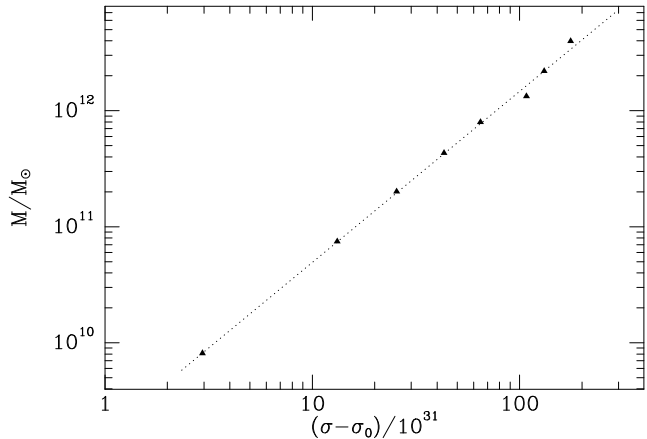}}
\caption{The function $M(\sigma)$ for the Hydra cluster estimated from the data
of \citet{dnm01}.  The
units of $\sigma$ are $ 10^{31} \cm^4\g^{-2/3}\s^{-2}$ and in these units
$\sigma_0=17$. The dotted line shows equation (\ref{mass}) with $\epsilon =1.48$.\label{hydras}}
\end{figure}

\section{The model}

Gas that is in stable hydrostatic equilibrium is stratified by specific
entropy $s$, such that $s$ never decreases as one moves up the gravitational
potential. Heating by an AGN disturbs this equilibrium by producing bubbles
of high-entropy gas near the bottom of the well. On a dynamical timescale
each bubble rises through the well, shedding as it goes gas of the
appropriate value of $s$ to each isopotential shell.  Following this
rearrangement, the gas is in approximate hydrostatic equilibrium,
but slowly radiates entropy. After a finite time, the gas at the bottom of
the well, which initially has the least entropy and the highest density, has
radiated all its entropy. Its density is then extremely large and a further
outburst of the AGN is stimulated when some of this dense gas accretes
onto the central massive black hole. This new outburst starts the next cycle
of the system.

If $P$ denotes pressure, $\rho$ gas density and $\gamma$ the ratio of
the principal specific heats, then
$s=\hbox{constant}+\fracj32\kB\ln(\sigma)$, where $\sigma\equiv
P\rho^{-\gamma}$ will be called the `entropy index' and $\kB$ is the
Boltzmann constant.  Let $M(\sigma)$ be the gas mass with entropy
index less than $\sigma$; Fig.~\ref{hydras} shows an estimate of
$M(\sigma)$ for the Hydra cluster from the data of \citep[][hereafter
D2001]{dnm01} plotted such that a straight line corresponds to the
functional form \[ M(\sigma)=A(\sigma-\sigma_0)^\epsilon.
\label{mass}
\]
 Since the data fall near a line, we assume that, at least during some
phases in the evolution of the cluster gas, $M(\sigma)$ can be
adequately fitted by this functional form. For the moment this choice
is motivated by the observations of the Hydra cluster. We will show in
Section \ref{power} that, even if at some early epoch, e.g. after an
AGN activity phase, $\sigma = {\rm const.}$ throughout the cluster, a
power-law distribution of $M(\sigma)$ naturally arises due to the
effects of radiative cooling. The fit involves three free parameters,
$A$, $\epsilon $ and $\sigma_0$. The lowest entropy density index
found in the cluster gas is $\sigma_0$ at the cluster centre. It is
therefore intimately related to the time that must elapse before the
next outburst of the AGN. The total mass of the cluster gas is given
by
\[
M_{\rm tot} = A (\sigma _{\rm max} - \sigma_0)^\epsilon,
\label{A_def}
\]
where $\sigma _{\rm max}$ is the highest entropy index of the cluster
gas at the largest radius. This relation defines the constant $A$.

Given a functional form $M(\sigma)$ one can find the hydrostatic equilibrium
configuration of the gas in a given spherical gravitational potential
$\Phi(r)$ as follows. Eliminating $\rho$ from the equation of hydrostatic
equilibrium we have
 \[
\frac{\d P}{\d r} = - \left( \frac{P}{\sigma} \right) ^{1/\gamma} \frac{\d
\Phi}{\d r}.
\label{equi}
\]
 A second equation relating $P$ and $M$ is
\[
\frac{\d M}{\d r} = 4 \pi r^2 \rho = 4 \pi r^2 
\left( \frac{P}{\sigma} \right)^{1/\gamma}.
\label{no2}
\]
 For a given $\Phi (r)$ we obtain a model atmosphere  by solving the
coupled differential equations (\ref{equi}) and (\ref{no2}) from $r=0$
with the initial conditions $M(0)=0$ and $P(0)=P_0$, where $P_0$ is a
trial value. At some radius $R_{\rm out}$ the gas mass reaches the value
$M_{\rm tot}$. The central pressure $P_0$ is adjusted until the pressure at
$R_{\rm out}$ is equal to a specified value $P_\infty$. 

It is instructive to recast equation (\ref{equi})  in terms of the
gas temperature, $T$. Assuming the cluster gas to be ideal, we can
eliminate $P$ from equation (\ref{equi}) and find
\[
\gamma \frac{\d\ln T}{\d r} = \frac{\d\ln\sigma}{\d r} - \left(
\gamma -1 \right) \frac{\mu m_{\rm p}}{\kB T} \frac{\d\Phi}{\d r},
\label{temp1}
\]
 where $\mu$ is the mean molecular weight. Since the second term on
the right side of this equation is intrinsically positive, the
indications from observations that $\d T/\d r>0$ at most radii
requires $\d\ln \sigma/\d r$ to be significantly greater than
zero. Now \[ {\d\ln\sigma\over\d r}= 4 \pi {\d\ln\sigma\over\d M}\rho
r^2,
\]
 so $\d\ln T/\d r$ can be positive near the centre only if $\d
M/\d\ln\sigma$ vanishes rapidly as $M\to0$. For our initial assumption
for $M(\sigma)$, equation (\ref{mass}), this requirement
translates to 
\[
{1\over\epsilon}\left(M\over M_{\rm tot}\right)^{1/\epsilon}
>{\gamma-1\over\sigma_{\rm max}/\sigma_0-1}\left({M\over4\pi r^2\rho}\right)
\left({\mu
m_{\rm p}\over\kB T}\right){\d\Phi\over\d r}.
\label{unequ}\]
 As $r\to0$, $\d\Phi/\d r\to\hbox{constant}$ in a Hernquist model and $M/
r^2\rho\sim r$ for any plausible density profile. Hence the right side of
this inequality vanishes with $r$ when $T$ tends to a constant, and as
$r^{1-\delta}$ if $T\sim r^\delta$ near the centre.
If the left side is to vanish more slowly,
we require $\epsilon >3/(1-\delta)$ where $\rho\sim\hbox{constant}$ and thus
$M\sim r^3$.

\subsection{Comparison with Hydra cluster}

In this section we calculate model atmospheres for the gas in galaxy
clusters. We adjust the model parameters to provide a fit to the
density and temperature distributions of the Hydra cluster as inferred
from X-ray observations (D2001). We do not attempt a formal fit of the
data as this would require folding our model results with the
telescope response. We are only interested in a rough constraint on
our model parameters to find reasonable values. Thus varying the model
parameters and fitting by eye is sufficient. From Fig.~\ref{hydras} we
find that for the Hydra cluster $\epsilon\sim1.5$ in equation
(\ref{mass}) and $\sigma _0 = 17 \times
10^{31}\cm^4\g^{-2/3}\s^{-2}$. Out to the limit of the data at $R_{\rm
out} \sim 230\kpc$, the ratio $\sigma _{\rm max} / \sigma_0\sim13.5$.
We assume that the gravitational potential has the form of the
potential of an NFW dark-matter halo \citep{nfw96}.  Then there are
four free parameters to determine by fitting the observational data:
the gas pressure at the centre of the cluster, $P_0$, the gas pressure
at $R_{\rm out}$, $P_{\infty}$, the scale length of the NFW profile,
$r_s$, and the overdensity of the dark matter halo compared to the
critical density at the redshift of the cluster, $\delta _c$.

\begin{figure}
\centerline{\psfig{file=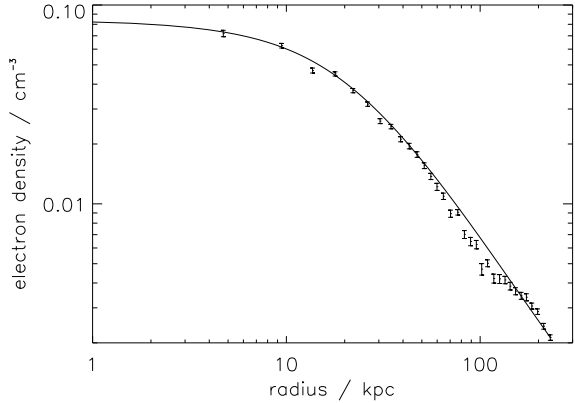,width=\hsize}}
\caption{The fit between the unevolved model (curve) and the density in the Hydra
cluster inferred from X-ray observations. Data points and errors from
D2001.\label{hydrad}}
\end{figure}

\begin{figure}
\centerline{\psfig{file=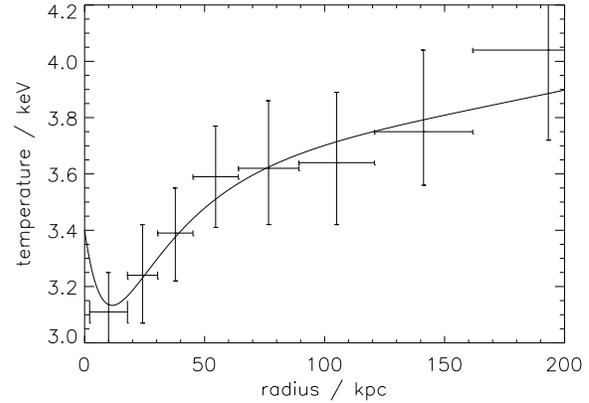,width=\hsize}}
\caption{Comparison of the initial model with temperature of the gas
in the Hydra cluster inferred from X-ray observations. Data points
show the temperatures derived by \citet{dnm01}. Vertical error bars
are the 90\% confidence limits and horizontal bars indicate the region
within the cluster over which the temperature was averaged. The solid
line shows the model.\label{hydrat}}
\end{figure}

Figs~\ref{hydrad} and \ref{hydrat} show comparisons of the model with
the data. The good fit is achieved for $P_0 = 8.8 \times
10^{-10}\erg\cm^{-3}$ and $P_{\infty} = 2.6 \times
10^{-11}\erg\cm^{-3}$. The length scale of the NFW profile is $r_s =
79.3\kpc$ and the overdensity of the dark matter halo $\delta_c = 8.5
\times 10^4$ for our chosen cosmology. Both values are consistent with
the findings of D2001. From these parameters we find the gas density
in the cluster centre is $\rho _0 = 1.7 \times 10^{-25}\g\cm^{-3}$,
which corresponds to an electron density of $0.085\cm^{-3}$. The gas
temperature at the centre is $\kB T_0=3.4\keV$. The mass enclosed
within $R_{\rm out} = 230\kpc$ is $M_{\rm tot} = 5.8 \times
10^{12}\msun$.

Since in our model $\rho$ tends to a constant at small $r$, the
inequality (\ref{unequ}) cannot be satisfied with $\epsilon = 1.5$. The
upturn in the temperature at small $r$ is a consequence of this fact.

\subsection{Cooling}

\begin{figure}
\centerline{\psfig{file=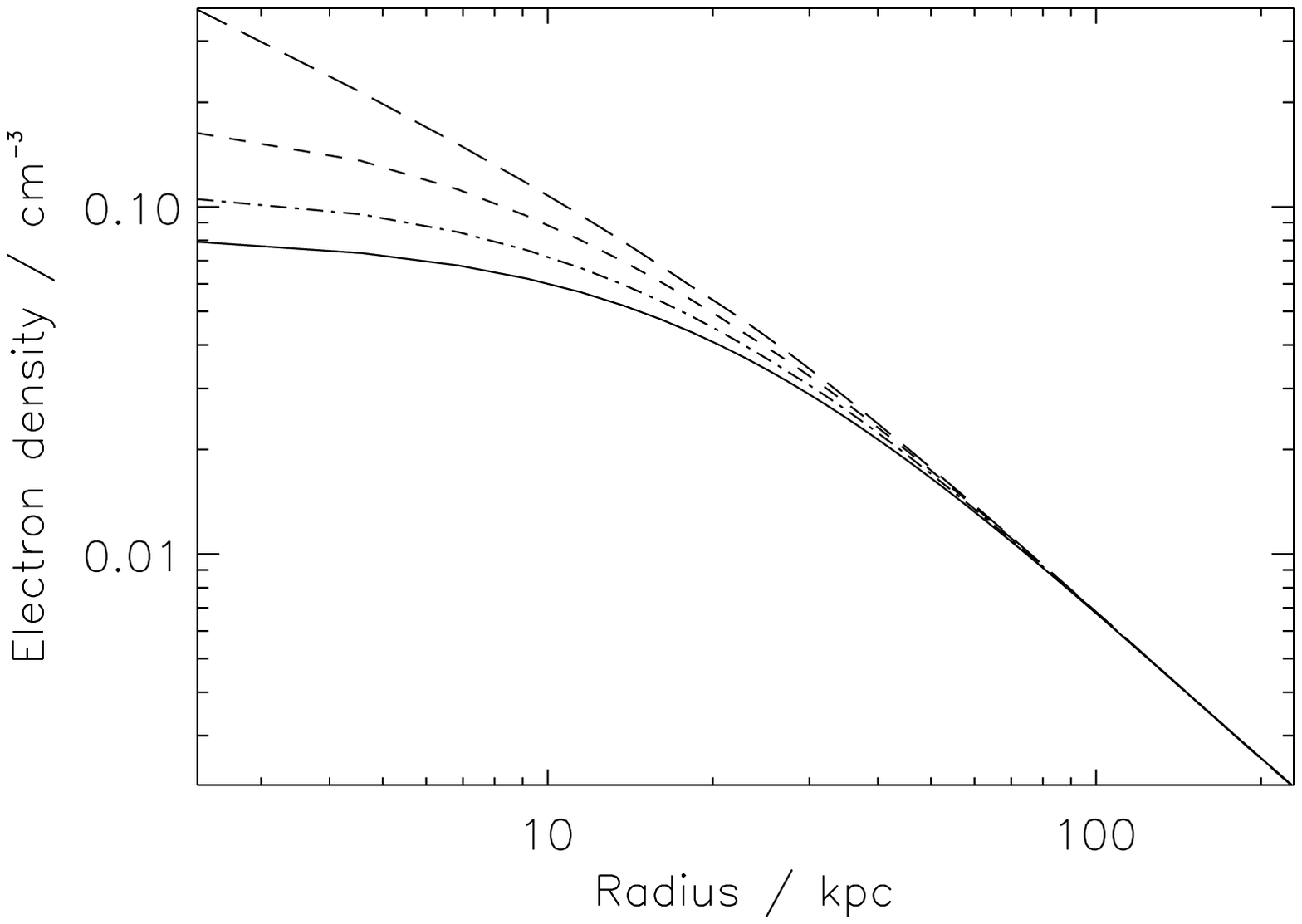,width=.95\hsize}}
\centerline{\psfig{file=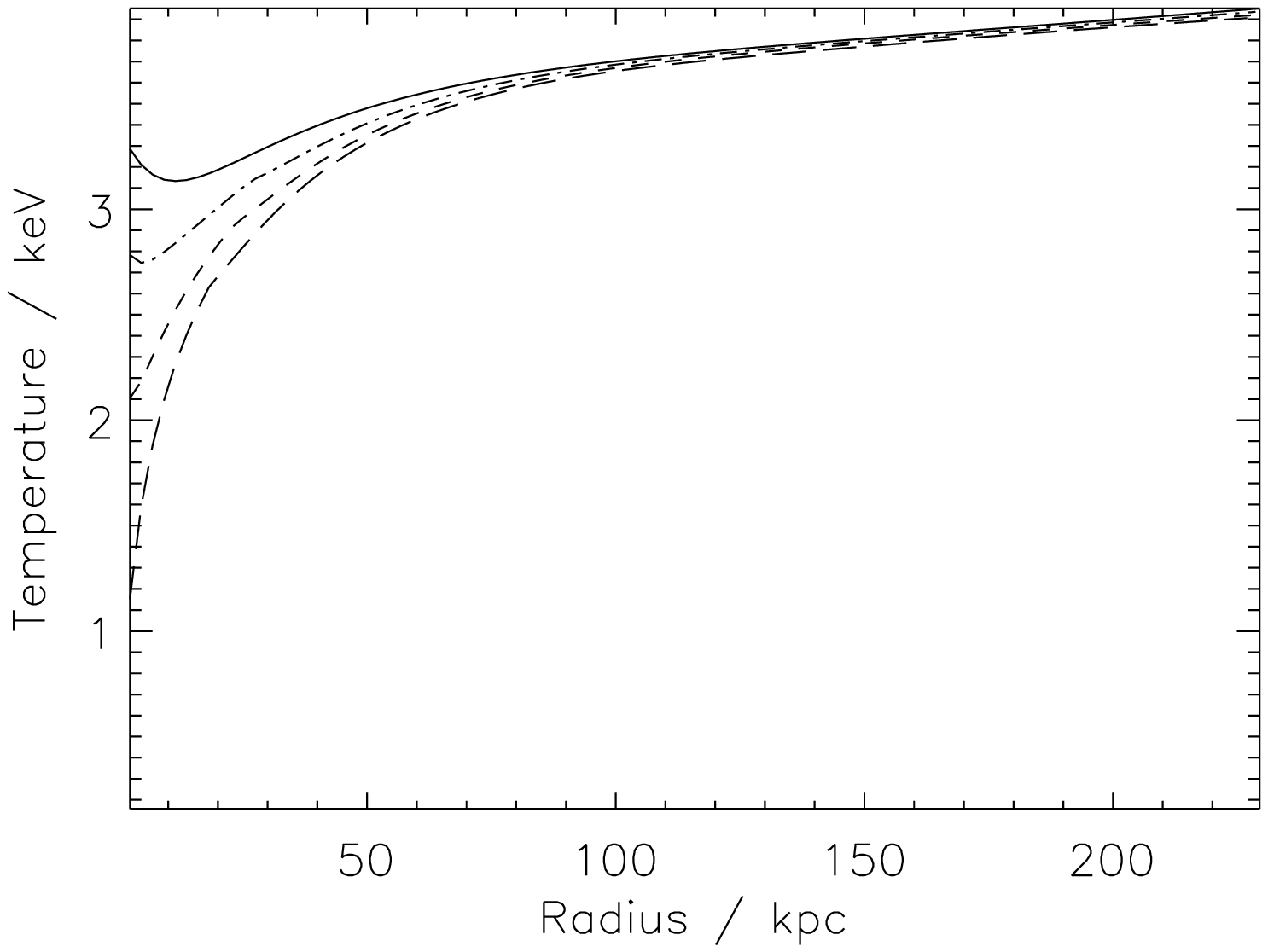,width=.95\hsize}}
\caption{Density and temperature profiles for the Hydra cluster at the
present time (full curves), after $100\Myr$ (dot-dashed curves), $200\Myr$
(short-dashed curves), and $284\Myr$ (long-dashed curves).\label{HydranT}}
\end{figure}

If $\Lambda(T)$ is the usual cooling function, the
rate at which specific entropy is radiated is 
 \[\label{cooleq}
\dot s(r,t)=\fracj32\kB{\dot\sigma\over\sigma}=-{\Lambda(T)\over T}n,
\label{sigtime}
\]
 where the electron density is
\[
n_\e(r)=0.85{\rho(r)\over m_p}.
\]
 From Fig.~9-9 of \citet{bt87} we calculate the cooling
function  $\Lambda(T)$ appropriate to metallicity $Z=0.4 Z_\odot$ (D2001).
Then by integrating equation (\ref{cooleq}) at each value of $r$ for a
small time interval $\delta t$, we can update the function $\sigma(M)$
that is initially specified by equation (\ref{mass}) 
 \[
\sigma(M)\to\sigma-\fracj23\sigma n{\Lambda(T)\over\kB T}\delta t,
\]
 where the quantities are evaluated at the radius that at time $t$
contains mass $M$. Assuming $P_{\infty} = {\rm
const.}$, we then use equations (\ref{equi}) and (\ref{no2}) to
reconstruct the density profile at time $t+\delta t$, and so on. Note
that $R_{\rm out}$ will decrease as the cluster gas radiates 
energy, and the material outside $R_{\rm out}$ slightly compresses the
cluster. Fig.~\ref{HydranT} shows the resulting predictions
for the evolution of the density and temperature profiles of the Hydra
cluster.

The value of the central temperature halves, and the central density doubles
in roughly $240\Myr$. The central temperature then drops rapidly to zero in
the following $44\Myr$. We assume that at this time the central AGN
erupts and restores the density and temperature profiles to distributions
similar to those currently measured.

\begin{figure}
\psfig{file=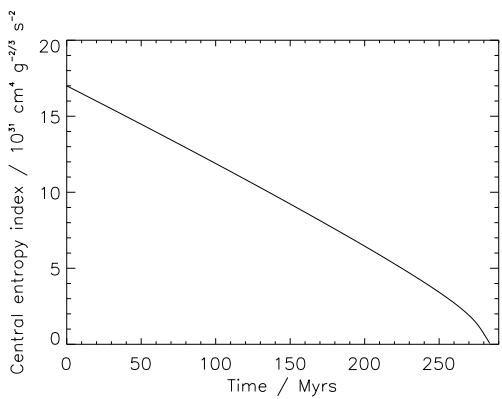,width=.9\hsize}
\caption{The time-evolution of the central entropy index
$\sigma_0$.\label{cenentrop}} 
\end{figure}

Fig.~\ref{cenentrop} shows the time-evolution of the central entropy
index $\sigma_0$. To a very good approximation, it falls linearly in
time between $t=0$ and $t=260\Myr$. For temperatures above about
$2.6$\,keV our cooling function is dominated by bremsstrahlung,
i.e. $\Lambda (T) \propto \sqrt{T}$. In this regime equation
(\ref{sigtime}) implies
\[
\dot{\sigma_0} \propto P_0^{2/5} \sigma _0^{1/10}.
\label{linear}
\]
The very weak dependence of $\dot\sigma_0$ on either $P_0$ or $\sigma _0$ leads to the
very nearly linear time dependence of $\sigma _0$. Even for
temperatures below $2.6\keV$, $\Lambda (T)$ and therefore
$\dot{\sigma_0}$ are not strong functions of temperature,
so the approximately linear time-evolution of $\sigma _0$ holds at nearly
all times. Of course, equation (\ref{linear})
also holds for material further out in the cluster. A linear
relation between $t$ and $\sigma$ can thus be used for an accurate estimate of the time to
elapse before the next nuclear eruption.

\begin{figure}
\psfig{file=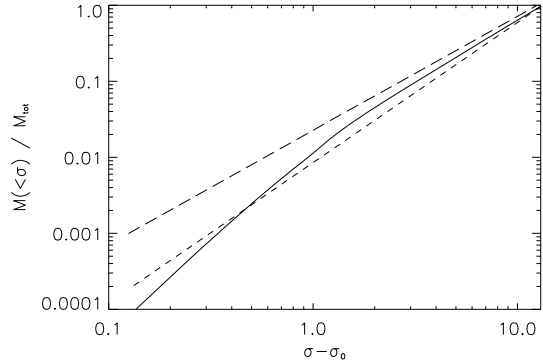,width=.9\hsize}
\caption{Cumulative mass of the cluster gas as a function of the
entropy index. Entropy index is shown in units of the current minimum
entropy index of the Hydra cluster. The long-dashed line shows the
initial power-law of equation (\ref{mass}). The solid line shows
$M(\sigma)$ at $t=284\Myr$. The short-dashed line represents the
power-law giving the best fit to the solid line. \label{msrel}}
\end{figure}

Figure \ref{msrel} shows the cumulative mass of the cluster gas as a
function of $\sigma-\sigma_0$, where $\sigma_0(t)$ is the current lower bound on the entropy index.
 Only at times later than about $230\Myr$ is there a
significant deviation from the power-law form of equation
(\ref{mass}).
 In fact, even at $t=284\Myr$ a power-law with
exponent $\epsilon = 1.8$ approximates $M(<\sigma)$ to within a factor
2. For gas with an entropy index greater than $2 \sigma _0$ a
power-law provides an excellent fit throughout the cluster evolution.

\begin{figure}
\psfig{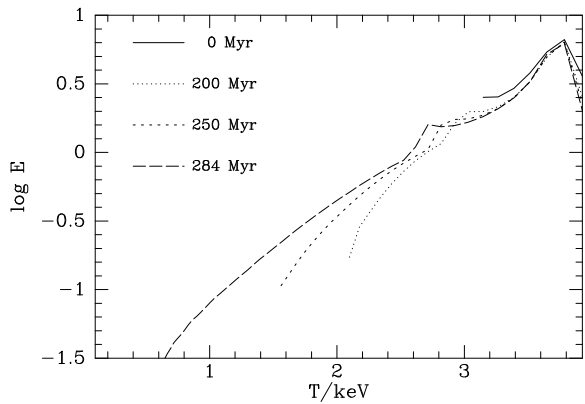}
\caption{The emission measure of the cluster gas at four different
times. \label{EMfig}}
\end{figure}

The emission measure distribution  is defined to be
\[
E(t,T)=\int\d V\,\delta(T'-T)n_{\rm e} ^2,
\]
 where $T'$ is the temperature of the gas in the volume element $\d
V$. Figure \ref{EMfig} shows the run of $E$ with $T$ at four
times. The evolution of $E(t,T)$ is dominated by the growth of a tail
towards low $T$ due to radiative cooling.  Gas at $T<1\keV$ appears
only at the very end of the evolution. Even gas with $T<2\keV$ appears
only at times later than about $245\Myr$.  We will show in Section
\ref{tmin} that this result explains the lack of observations of very
cold gas in galaxy clusters \citep[e.g.][]{bmc02}. The secondary peak
at temperatures below the main peak is caused by the break in the
density distribution at around 30\,kpc (see Figure \ref{HydranT}).

\subsection{The entropy index-mass distribution}
\label{power}

Inspired by observational data for the Hydra cluster, we have adopted an
initial entropy index-mass distribution, $M(\sigma)$, of the form given in
equation (\ref{mass}). We have shown that during the subsequent cooling of
the cluster gas this power-law form is preserved (see Fig.~\ref{msrel}).
Thus, any cluster for which $M(\sigma)$ is given by equation (\ref{mass})
at one point in its
evolution, will retain this form of $M(\sigma)$ also at later times. 
It is natural to ask whether the relationship between $M$ and
$\sigma-\sigma_0$ evolves towards  the power-law form if a cluster is
initially configured in a different way. To answer this question rigorously,
one would have to follow a cluster's evolution from infinitely many initial
conditions. A configuration of special interest is the adiabatic one in
which all the gas has the same entropy index $\sigma_0$: 
convection, whether it is caused by AGN activity or infall, drives the gas
towards this configuration.

\begin{figure}
\psfig{file=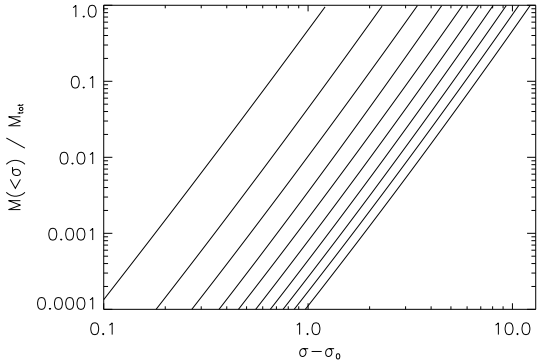,width=.9\hsize}
\caption{Cumulative mass as a function of entropy index for the
cluster with an initially constant entropy index throughout.  Entropy
index is shown in units of the current minimum entropy index of the
Hydra cluster. The lines show the distribution at intervals of 1\,Gyr
from 1\,Gyr (top) to 10\,Gyr (bottom) after the start of the
calculation.
\label{straight}}
\end{figure}

Fig.~\ref{straight} shows the results of using the iterative procedure
described in the previous section to follow the radiative evolution of an
initially adiabatic cluster atmosphere.  The parameters are set to the same
values as for our study of the Hydra cluster except that the initial entropy
index of all the gas in the cluster is $3.7 \times
10^{33}\cm^4\g^{-2/3}\s^{-2}$, 22 times larger than the current minimum
entropy index in Hydra, and that the initial pressure of the gas at the
centre of the cluster is $1.7 \times 10^{-10}$\,erg\,cm$^{-3}$, a fifth of
the central pressure in Hydra.  These values were chosen so that the cluster
gas evolves into a state that roughly resembles the current state of the
Hydra cluster. 

From top to bottom, the curves in Fig.~\ref{straight} show the $M(\sigma)$ relation in the
cluster at $t=1,2,\ldots,10\Gyr$ after the start of the calculation. All
distributions are virtually indistinguishable from the power-law form
of equation (\ref{mass}) with an appropriate choice of $\epsilon$. In
fact, the power-law form of $M(\sigma)$ is effectively established as early
as  $\sim0.1\Gyr$  (not shown in the Figure). At 10\,Gyr the
minimum entropy index of the cluster gas is equal to the currently measured
value of $\sigma_0$ in Hydra. The
entropy index also spans a range similar to that of the gas in the
Hydra cluster. Only the exponent of the power-law distribution,
$\epsilon = 3.4$, is considerably steeper. Starting from an initially
uniform entropy index distribution, the precise value of $\epsilon$
that results at later times depends sensitively on the chosen initial
conditions. Also, as before we make the assumption that the pressure
at the outer boundary of the cluster, $P_{\infty}$, is constant during
the entire evolution of the cluster. Considering the long timescales
involved and the significant contraction of the outer radius due to the
cooling process over this time, it is unlikely that $P_{\infty}$
really remains constant. Any evolution of the pressure at the outer
boundary will influence the value of $\epsilon$ at later times. A
more detailed study of these effects is beyond the scope of this
paper.

Thus a cluster that initially contains gas with a uniform entropy index,
almost instantaneously develops a power-law distribution for $M(\sigma)$.
Once the power-law is established, it persists, with changing exponent
$\epsilon$, until the gas at the centre of the cluster cools to very low
temperatures.  These results suggests that 
the observed power-law distribution $M(\sigma)$ of the Hydra
cluster would appear to be a generic rather than an exceptional
configuration.

\section{The creation and loss of entropy}

We now estimate the rate at which the AGN creates entropy in the ICM and
compare it with the rate at which entropy is radiated by the ICM. We
idealize the blowing of a bubble as a process in which $N_1$ particles of
the ICM are irreversibly heated to a high temperature $T_1$ from the
temperature $T_3$ at the base of the cooling flow. The heated bubble expands
as it is heated and does work on its environment. The interface between the
bubble and the ambient medium is Rayleigh-Taylor unstable so it will
dissolve on a dynamical time. From the sharpness of the edges of observed
bubbles it thus follows that they inflate supersonically. Since material
just outside bubbles appears not to be on a high adiabat
\citep{fsetacijo00}, we infer that the expansion is only mildly supersonic.
The entropy generation by weak shocks is small (see appendix), and we
neglect it.

The entropy of a non-degenerate ideal monatomic gas of $N$ particles that
occupies volume $V$ at temperature $T$ is
 \[\label{Eofgas}
S=N\kB\left\{\fracj52+
\ln\left[{V\over N}\left({2\pi m\kB T\over h^2}\right)^{3/2}\right]\right\}.
\]
 Hence, when a bubble containing $N_1$ particles is heated at constant
pressure from $T_3$ to $T_1$, the entropy created is
 \[\label{Sinfl}
S_{\rm infl}=\fracj52N_1\kB\ln(T_1/T_3).
\]

Additional entropy is created as the bubble rises and  mixes in with the
ambient medium. Since most of the cluster gas falls within quite a small
range in temperature,
we simplify the equations by supposing that the bubble mixes with material
that all has temperature $T_2$. Let the
bubble mix with material that contains $N_2$ particles and assume for
simplicity that the mixing occurs at constant pressure. Then from
eq.~(\ref{Eofgas}) it is straightforward to show that the entropy of
mixing is
\begin{eqnarray}\label{Smix}
S_{\rm mix}&=&
\fracj12N_1\kB\ln\left({1+N_2T_2/(N_1T_1)\over1+N_2/N_1}\right)\nonumber\\
&&+\fracj12N_2\kB\ln\left({1+N_1T_1/(N_2T_2)\over1+N_1/N_2}\right)
\end{eqnarray}
 Adding eqs (\ref{Sinfl}) and (\ref{Smix}) we get the total entropy created
by a bubble. As an illustrative example, if $T_1=10 T_2$, $T_3=T_2/10$ and
$N_1=N_2/10$, then $S_{\rm infl}=11.5\kB N_1$ and $S_{\rm mix}=2.14\kB N_1$.
The relative contribution to the total entropy production from $S_{\rm mix}$
increases with $N_2/N_1$; for example with $T_1=10T_2$ again but
$N_2=100N_1$, $S_{\rm infl}$ is unchanged while $S_{\rm mix}=3.2\kB N_1$.

\begin{figure}
\centerline{\psfig{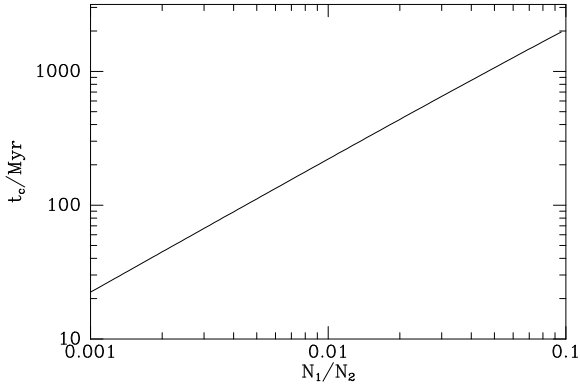}}
\caption{Time $t_{\rm cool}$ required to radiate entropy generated by
processing a mass fraction $N_1/N_2$ of the cluster plasma through bubbles
heated to 10 times the ambient temperature.\label{tcoolfig}}
\end{figure}

We may estimate the fraction of the plasma that is processed through bubbles
during an outburst by finding the time taken for gas of density $N_2/V_2$ at
temperature $T_2$ to radiate the entropy created in an outburst, where $V_2$
is the volume of the gas. The rate of entropy radiation is
 \[
\dot S={\Lambda(T_2)(0.52N_2/V_2)^2V_2\over T_2},
\]
 where $0.52$ is the conversion factor between total particle density
$N_2/V_2$ and electron density.
The time to radiate the entropy created by a bubble is thus
 \begin{eqnarray}
t_c={S\over\dot S}&=&
{N_1\over N_2}\bigg\{
5\ln\left[{T_1\over T_3}\right] +
\ln\left[{1+N_2T_2/(N_1T_1)\over1+N_2/N_1}\right]\nonumber\\
&&+{N_2\over N_1}\ln\left[{1+N_1T_1/(N_2T_2)\over1+N_1/N_2}\right]\bigg\}\tau,
\end{eqnarray}
 where
\[
\tau\equiv{\kB T_2\over0.55\Lambda(T_2)(N_2/V_2)}.
\]
 Fig.~\ref{tcoolfig} shows $t_c$ as a function of bubble mass
fraction $N_1/N_2$ for $T_1=10 T_2$, $T_3=T_2/10$ with
$N_2/V_2=0.02\cm^{-3}$ and $T_2=3\times10^7\,$K as numbers typical of the
central part of Hydra A (D2001). A bubble mass fraction $\sim1$
percent yields cooling times of the expected order, $200\Myr$.

\section{Distribution over $T_{\rm min}$}
\label{tmin}

In data taken with {\it Chandra\/} and {\it XMM/Newton\/} it is found that the plasma
in any given cluster is confined to temperatures $T_{\rm max}>T>T_{\rm
min}$, where typically $T_{\rm min}>10^7\K$ \citep[e.g.][and
references therein]{bmc02}. In terms of the model discussed here,
$T_{\rm max}$ and $T_{\rm min}$ are the largest and smallest
temperatures at which there is significant emission measure. We now
assume that the evolutionary time $t$ of clusters is uniformly
distributed in the interval $(0,t_{\rm cool})$, where $t_{\rm cool}$
is the time required for the central temperature of a freshly reheated
cluster to cool to zero.  With this assumption and the approximation
that we can neglect the evolution in the comoving density of clusters,
we can calculate the fraction of observed clusters for which $T_{\rm
min}$ lies in the range $(T+\d T,T)$. We assume that the clusters are
drawn at random from a flux-limited catalogue that was compiled with
an instrument of limited spatial resolution. Hence, we focus
exclusively on the total flux coming from plasma at a given
temperature, and ignore issues related to the spatial distribution of
the emission. This simplification is probably fairly realistic for
current data.

First we determine the emission measure distribution of our model as
shown in Fig.~\ref{EMfig}. Multiplying $E(t,T)$ by the cooling
function $\Lambda(T)$ and integrating over $T$ we obtain the cluster's
bolometric luminosity $L(t)$. Dividing $L$ by $4\pi D(z)^2$, where
$D(z)$ is the luminosity distance of an object at redshift $z$
\citep{cpt92}, we obtain the flux $S(z,t)$ of a cluster observed at
evolutionary stage $t$: \[ S(z,t)={1\over4\pi D(z)^2} \int_0^\infty\d
T\,\Lambda(T)E(t,T).
\]
 We are interested in the number of clusters with $S>S_{\rm min}$, so we
invert this relation to find the redshift $z_{\rm max}(t)$ out to which the
cluster has $S>S_{\rm min}$ at stage $t$. With our assumptions, the number of
clusters detected at evolutionary stages in $(t+\d t,t)$ is
proportional to $V[z_{\rm max}(t)]\,\d t$, where $V(z)$ is the comoving
volume to redshift $z$.

\begin{figure}
\centerline{\psfig{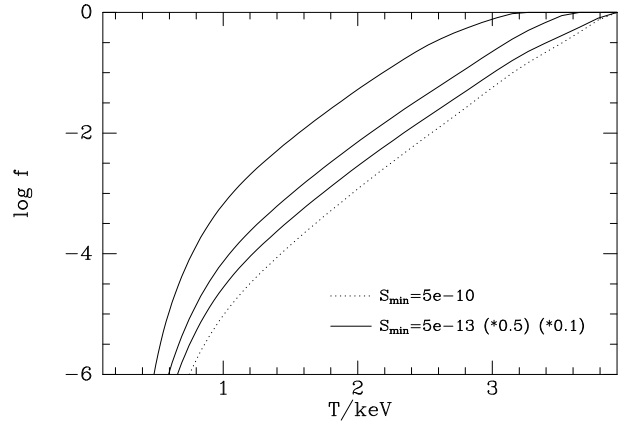}}
\caption{The fraction of detected clusters in which plasma cooler than
$T$ can be detected for two flux limits: $S_{\rm
min}=5\times10^{-10}\erg\cm^{-2}\s^{-1}$ (dotted curve) and $S_{\rm
min}=5\times10^{-13}\erg\cm^{-2}\s^{-1}$ (bottom full curve). The top full
curve shows the fraction of clusters in the fainter sample in which plasma
cooler than $T$ would be detected in a pointed observation ten times more
sensitive.
\label{fracfig}}
\end{figure}

Similarly, the flux from plasma cooler than $T_{\rm min}$ in a cluster at
redshift $z$ is
 \[
S(T_{\rm min},z,t)={1\over4\pi D(z)^2}
\int_0^{T_{\rm min}}\d T\,\Lambda(T)E(t,T),
\]
 which on inversion yields the maximum redshift $\hat z_{\rm max}(T_{\rm
min},t)$ out to which the cool plasma can be detected. Finally, the fraction
on observed clusters in which plasma cooler than $T_{\rm min}$ can be
detected is
 \[
f(T_{\rm min})={\int_0^{t_{\rm cool}}\d t\,V[\hat z_{\rm max}(T_{\rm min},t)]\over
\int_0^{t_{\rm cool}}\d t\,V[z_{\rm max}(t)]}.
\]
 The lower two curves in Fig.~\ref{fracfig} show $f(T_{\rm min})$ for
two values of the limiting flux $S_{\rm min}$, namely
$5\times10^{-13}$ and $5\times10^{-10}\erg\cm^{-2} \s^{-1}$
(dotted). For values of $S_{\rm min}$ smaller than about
$10^{-11}\erg\cm^{-2} \s^{-1}$ the distance to which even the warmest
plasma in Hydra can be seen is not large cosmologically, and $f(T_{\rm
min})$ becomes insensitive to $S_{\rm min}$.  At lower flux limits,
cosmological dimming becomes significant, especially for the warmest
plasma, and the fraction of clusters in which cooler plasma can be
detected rises slightly.

The lower two curves in Fig.~\ref{fracfig} predict that only
$\sim10^{-3}$ of clusters with bolometric luminosity in excess of
$S_{\rm min}$ will have a flux in excess of $S_{\rm min}$ from gas
cooler than $T\sim2\keV$. In practice when we examine a catalogue of
clusters discovered in a magnitude-limited survey, gas at $T<2\keV$
will have been detected in a fraction of clusters that exceeds
$10^{-3}$ because many clusters in the catalogue will have been the
subject of more sensitive pointed observations subsequent to their
discovery in the original survey. To quantify this effect, the upper
two full curves in Fig.~\ref{fracfig} show the fraction of clusters in
which gas at temperatures below $T$ gives rise to a flux in excess of
$0.5S_{\rm min}$ and $0.1S_{\rm min}$ (top curve). From these curves
we see that if the pointed observations are ten times as sensitive as
the survey limit, the fraction of clusters showing gas at $T<2\keV$
rises to 5.6 percent, while the fraction showing gas at $T<1\keV$ is
still only 0.04 percent.

\section{Distribution over $\sigma_0$}

\citet{nk91} pointed out that there is a gross discrepancy between the
predictions of gravity-driven hierarchical models of structure formation and
the statistics of X-ray astronomy. In particular, the theory of hierarchical
structure formation predicts that the comoving density of X-ray sources was
much higher at redshifts $z\gta1$ than is observed, and the cluster
mass--X-ray luminosity relation is less steep than that observed. Kaiser
pointed out that these conflicts would be eliminated if structure-formation
theory were extended to include heating of the intergalactic medium (IGM) by
massive stars and AGN prior to and at the epoch of galaxy formation. As
Kaiser pointed out, one should think of such heating as raising the adiabat
on which the IGM lies. \citet{lpc00} have attempted to determine the adiabat
of the IGM by measuring the entropy index $\sigma$ in galaxy groups and
clusters at a distance of 10 percent of each system's virial radius,
well outside the region in which radiative cooling is important. They define
an entropy index $s=\kB T/n_e^{2/3}$ with units of
$\keV\cm^2$. Since $n/n_{\rm e} = 21/11$, the conversion to our definition
of $\sigma$ is $ 1 \keV\cm^2 = 9.82 \times 10^{30}\cm^4\g^{-2/3}\s^{-2}$.
Their `entropy floor' of $s_{0.1}\sim 100\keV\cm^2$ is equal to the entropy
index of gas at a radius of $200\kpc$ in our model of the Hydra cluster.
This distance from the cluster centre corresponds to 10 percent of the
virial radius of Hydra. The agreement is not surprising considering that the
Hydra cluster was included in the study of \citet{lpc00}.

\begin{figure}
\centerline{\psfig{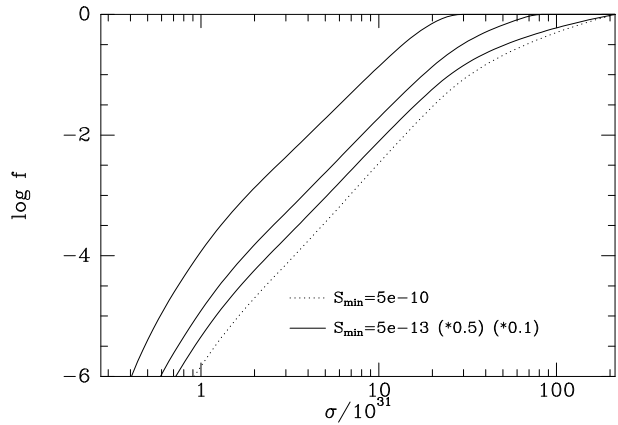}}
\caption{As Fig.~\ref{fracfig} but giving the fractions of clusters in which
plasma with entropy index smaller than $\sigma$ would be detectable.
The units of $\sigma$ are as in Fig.~\ref{hydras}. In these units
$\sigma=0.982\kB T n_e^{-2/3}$ with $T$ in keV and $n_e$ in $\!\cm^{-3}$.
\label{sminfig}}
\end{figure}

Lloyd-Davies et al.\ specifically excluded from consideration gas which they
considered dense enough to have cooled significantly because they wished to
probe the state of the IGM before clusters formed.  In the context of our
model of a cooling flow, it is interesting to ask whether they would have
encountered an effective entropy floor even if they had not excluded cooling
gas and had simply measured $\sigma_0$, the minimum entropy index at the
cluster centre. In Fig.~\ref{sminfig} we show the fraction of clusters that
have detectable emission from plasma with entropy index lower than $\sigma$
-- the calculation of this fraction proceeds in close analogy with that of
Section \ref{tmin}. We see that in a pointed observation ten times more
sensitive than the survey from which the targets were selected, only 1
percent of clusters would have detectable emission from plasma with
$\sigma<2\times10^{31}\cm^4\g^{-2/3}\s^{-2} \approx 2 \keV\cm^2$. Thus an
`entropy floor' with regard to the minimum entropy index of gas in galaxy
clusters should be detectable in X-ray observations of clusters that have
sufficient spatial resolution. This entropy floor is considerably lower than
that found by \citet{lpc00}. It is determined by the very rapid cooling of
the central regions of clusters, rather than any pre-heating of the gas
before the collapse of the cluster that may explain the entropy floor at
larger radii \citep{vb01,mbb02}.

\section{Discussion}

In the Hydra cluster we find that the distribution of mass in entropy index
$\sigma$ has a simple power-law form. We find that this form is to a good
approximation preserved when the cluster gas cools in hydrostatic
equilibrium. Furthermore, the activity of an AGN at the cluster centre
drives convective flows in the cluster gas. Continued convetion would
eventually lead to a cluster with gas of a constant entropy throughout its
entire volume. We show that starting from such a constant entropy
distribution, the power-law form observed in the Hydra cluster arises
naturally as the cluster gas cools. Thus the
discovery that $M(\sigma)$ for Hydra fits a power-law is no accident, but
would have been discovered if we observed Hydra at an earlier or a later
epoch.

We find that the central value of the entropy index, $\sigma_0$,
is a linear function of time until right up to the final cooling catastrophe
that provokes an outburst of the central galactic nucleus. These results
enable one to calculate the dynamics of a cooling flow with remarkable ease.

We estimate that in the absence of heat sources Hydra has about $280\Myr$ to
run before there is a central cooling catastrophe. This result places a lower
bound on the time between nuclear outbursts. The actual inter-outburst time
is likely to be  longer both because there is residual heating by
supernovae/galactic winds and a low-level AGN, and because we have no reason
to suppose that Hydra is in its immediate post-outburst state; it has
probably been cooling for some time.

In trying to understand a cooling flow it is useful to focus on the system's
entropy at least as much as on its energy. Between outbursts the story as
regards entropy is simple: it is being radiated at a rate that can be readily
calculated from the X-ray brightness profile. The story regarding energy is
much more complicated: the gas losses energy  radiatively but recovers some
 through the work done by both the surrounding IGM and
the intracluster gravitational field. Our premise is that the nuclear
outburst that is provoked by each cooling catastrophe restores the cluster
gas to essentially the same state that it had immediately after the previous
outburst. This premise requires that the entropy created by irreversible
processes during each outburst is equal to the entropy radiated between
outbursts.

We assume that the restructuring of the cluster gas during an outburst is
effected by bubbles of plasma. We calculate both the entropy created when a
bubble is inflated, and that created as it mixes in with, and heats, the
bulk of the intracluster gas. The relative sizes of these two entropy
sources is a weak function of the mass fraction that passes through bubbles
during an outburst, but is typically of order 3--5. If a few percent of the
cluster gas passes through bubbles during an outburst, the time between
outbursts is a few hundred Myr, as the observations suggest.

When gas cools from temperatures $\gta3\keV$ at which bremsstrahlung is the
dominant cooling process, it spends only a very small fraction of the total
cooling time at $T<1\keV$. In the case of cluster gas this effect is
magnified by the large density contrast between the cluster centre and the
half-mass radius. Consequently, not only does gas at $T<1\keV$ exist only
for a very small fraction of the inter-outburst time, but it is confined to
an extremely small fraction of the total volume. We have used our simple
models of cooling flows to calculate the fraction of clusters in a
magnitude-limited sample in which gas cooler than temperature $T$ would be
detectable. We find that without pointed observations gas cooler than
$2\keV$ would be found in only $\sim10^{-3}$ of clusters; with pointed
observations ten times more sensitive than the discovery survey, such cool
gas would be detected in $\sim4$ percent of clusters.
Thus, we should not be surprised to find that gas cooler than $\sim1\keV$
has not been detected in clusters observed by {\it XMM\/} and {\it Chandra}.

Similarly, we predict that pointed observations of a complete sample
of clusters would detect gas at entropies below $\sigma\sim2\keV\cm^2$
in only $\sim1$ percent of clusters because low-entropy gas appears
only fleetingly and in small volumes. This lower limit on the entropy
is at a value of $\sigma$ that is a factor $\sim50$ lower than the
`entropy floor' claimed for intergalactic gas from X-rays studies of
low-mass clusters at radii well outside the region where radiative
cooling is important.  However, with the spectral and spatial
resolution of {\it XMM\/} and {\it Chandra} the lower entropy limit
should be detectable in samples of galaxy clusters.

\section*{Acknowledgments}
We are grateful to L.P.\ David for supplying  data for the Hydra cluster in
digital form. We thank the anonymous referee for helpful comments.

\begin{figure}
\centerline{\psfig{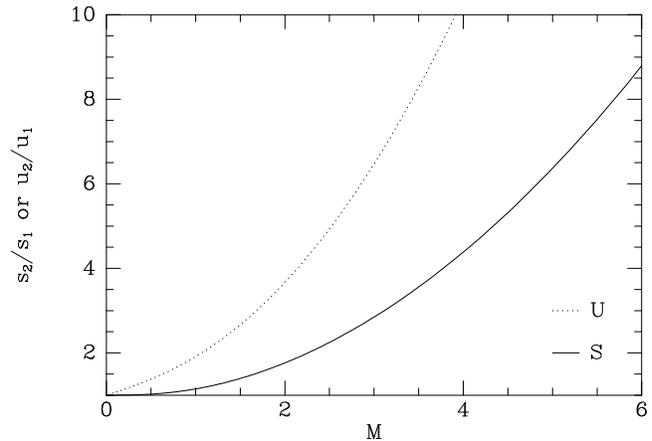}}
\caption{Entropy change across a shock.\label{entropfig}}
\end{figure}

\section*{Appendix: entropy generation in shocks}

A major source of entropy production is shock fronts. Consider a
perpendicular shock front in which an ideal monatomic gas with specific
entropy $s_1$, specific internal energy $u_1$ and speed $v_1$ is converted
into material in which the same quantities are denoted $s_2$, $u_2$ and
$v_2$. Then conservation of mass, momentum and energy across the shock front
yield
 \[
{s_1\over s_2}={r_1\over r_2}\left({5r_2/3+1/2\over5r_1/3+1/2}\right),
\]
 where the $r_i\equiv u_i/v_i^2$ are the two roots of the quadratic equation
\[
{(2r/3+1)^2\over5r/3+1/2}=\hbox{constant}.
\]
 Fig.~\ref{entropfig} shows the changes in $u$ and $s$ across a shock
front as a function of the Mach number\footnote{Our definition of $M$
differs significantly from a common one that involves the upstream
speed relative to the shock. With our definition, in the limit of a
sound wave $M\to0$, rather than $M\to1$ as in the case of the
alternative definition.} of the shock, $M=(v_1-v_2)/c_{\rm s}$. The
very slow rise of the curve for $s_2/s_1$ at small values of $M$
implies that weak shocks are very ineffective entropy generators.
This is also shown in \citet{ll87}. The much steeper rise in the
curve for $u_2/u_1$ shows that weak shocks do thermalize energy
effectively, but since they do so without significant entropy
increase, the original bulk kinetic energy can be largely recovered by
subsequent adiabatic expansion.

\end{document}